\documentclass[a4paper]{article}

\usepackage{graphicx,amsmath}
\def\FRAC#1#2{\leavevmode\kern-.1em
 \raise.5ex\hbox{\the\scriptfont0 #1}\kern-.1em
  /\kern-.15em\lower.32ex\hbox{\the\scriptfont0 #2}}
\begin{document}

\title
{\bf Electron trajectory in the hydrogen atom}

\author
{
Yoshio {\sc Nishiyama}\\
\small Department of Science Education,
 Faculty of Education and Human Sciences \\
\small Yokohama National University, Yokohama,
  240-8501 Japan
}
\date{ nisiyama@ed.ynu.ac.jp}
\sloppy
\maketitle

\renewcommand{\abstractname}{}
\begin{abstract}
A trajectory in the Schr\"odinger wave for an electron in an
 attractive Coulomb potential with the dynamical behavior is
 proposed and illustrated for a scattering and a bound state.
 The  scattering cross section derived from the trajectories
 is almost exactly equal to that from the usual wave theory.
The statistical nature of the result is examined.
  The period of one cycle in the bound state is
   exactly equal to that of the corresponding classical motion.
\end{abstract}

\section{Introduction}
Nowadays, we can move an atom to put on the arbitrary positions
 on metal.
We can measure whether one electron adhered to the capacitor.
There is no uncertainty to operate one atom.
We also have ascertained that there is no indeterminacy
on determining a submicron diameter by a longer wavelength
laser precisely.~\cite{OptRev01}
It would be one of the urgent requisites to make a dynamical equation
 of motion of an atomic scale `particle' without any contradiction
 to the wave theory.

The trajectory of the harmonic oscillator in the wave equation
 has been discussed.~\cite{Nishiyama7}
Since the system treats only the bound state, the statistical nature, one of
the important characteristics, in quantum theory can not be seen clearly.
In the present paper, by using the method of
 the mcf~\cite{Nishiyama7,JOSAA95},
 a trajectory of an electron in an attractive Coulomb potential, like
  the hydrogen atom in a state with a positive or negative energy, is proposed.
On discussing the cross section for the scattering state, the statistical
but in principle determinate nature like classical mechanics
 will be seen as an illustration.

In  section \ref{sec:QDWF} the dynamical theory that would lead to
 a trajectory is described.
 The  traveling waves and usual stationary wave functions are discussed.
In section \ref{sec:SCAT} the trajectory of an electron in the
scattering state by the Coulomb potential is investigated.
In sections \ref{sec:crossect} and \ref{sec:flux} the cross section and
the flux of the beam of particles are discussed.
  In section \ref{sec:HYDRSCH} the trajectory of an electron
 in the hydrogen atom in a  bound state is analyzed and illustrated.
Conclusion and remarks on the wave functions are  given
 in section \ref{sec:Concl}.

\section{Dynamics and wave function}\label{sec:QDWF}
 The dynamics that leads to the mode trajectory of an electron
 in an attractive Coulomb potential with a charge $e ( > 0)$ is
  summarized.~\cite{JOSAA95}
 The wave function  $\Psi$ describing the motion of an electron
  satisfies the Schr\"odinger equation
\begin{eqnarray}
 i\hbar {\partial \Psi ({\bf r},t) \over \partial t}
   = \left(  - {\hbar^2 \over 2m} \triangle  - {e^2 \over r} \right)
    \Psi ({\bf r},t),
   \label{eq:Schroe}
\end{eqnarray}
 where constant $m$ or $-e$ is electron mass or charge, respectively.

The equation is assumed to be separable in variables $t, x_1, x_2$
 and $x_3 $.
Let the wave function be
\begin{eqnarray}
 \Psi ({\bf r},t) = e^{-iEt/\hbar} \Phi_1(x_1,E,\alpha)
    \Phi_2(x_2,\alpha ,\beta ) \Phi_3(x_3,\beta ),
\end{eqnarray}
 where  $E, \alpha$ and $\beta $ are constants of separation, and $E$ is
 assumed to be the energy of the system.
 These constants should be called mode parameters.
 The wave function of the form
\begin{eqnarray}
   \Phi_j(x_j) = \sqrt{P_j (x_j)} \exp \{ iW_j (x_j) \}, \qquad
     j = 1, 2, 3,
                                    \label{eq:Wjxj}
\end{eqnarray}
 is sought, where functions $P_j$'s and $W_j$'s are real.
This should be called a traveling wave.
 Let functions $W_j$'s  satisfy the condition that
  in each classical region of $x_j$ for $j = 1,2,3$
 \begin{eqnarray}
      W_j(x_j) \simeq  W_j(x_j)_{\rm cl},
  \end{eqnarray}
 where the sum of them
 \begin{eqnarray}
      W_{\rm cl} = \sum_{j = 1}^3 W_j(x_j)_{\rm cl}
  \end{eqnarray}
 is the Hamilton characteristic function of the Hamilton-Jacobi
 equation in classical mechanics.~\cite{Goldstein}
The classical region stands for the domain in which the chracteristic
 function holds true.

 If  $W_j$'s are found uniquely, the sum of them
 \begin{eqnarray}
 W(x_1, x_2, x_3, E, \alpha , \beta ) = W_1(x_1,E, \alpha )
  + W_2(x_2, \alpha ,\beta ) + W_3(x_3, \beta )
                     \label{eq:Wmcf}
 \end{eqnarray}
 is named the mode characteristic function (mcf) for the system.

 The equations of motion for the electron are assumed
 \begin{eqnarray}
 {\partial W \over \partial E} = {1 \over \hbar}(t - t_0),  \qquad
  {\partial W \over \partial \alpha } = c_{\alpha}, \qquad
    {\partial W \over \partial \beta } = c_{\beta },
                  \label{eq:MotionS}
 \end{eqnarray}
where $ t_0, c_{\alpha}$ and $c_{\beta}$ are constants (independent of $t$)
 that are determined by initial conditions for the system.
 Variable  $t$ is considered  the dynamical time for the system.
 The trajectory derived from Eqs.~\eqref{eq:MotionS}
 should be  called the mode trajectory.

The square integrable wave function is obtained as follows.
Let the endpoints of the $x_j$ coordinate be $a_j$ and
 $b_j$.
 The electron moves in the region $a_j \le x_j \le b_j$.
Let it start from a point $x_{j0}$ to the increasing $x_j$ direction
and the mcf in the coordinate space be $W_j(x_j)$.
The traveling wave associated with the motion reversing from $b_j$
to $a_j$ should be assumed to be given by
 $ \sqrt{\rho_j(x_j)} \exp [ i\{ -W_j(x_j) + 2W_j(b_j) \} ]$,
 which is also the solution of Eq.~\eqref{eq:Schroe}.

The wave  observed at $x_j (\ge x_{j0})$ should be the superposition
 of the traveling waves associated with the alternating motion
  of the electron~\cite{GORDON}
 \begin{eqnarray}
     \Phi_j(x_j) - \Phi_j^*(x_j)e^{ i 2W_j(b_j) }.
                    \label{eq:xjbj}
\end{eqnarray}
 This is finite, zero, at $b_j$.
  If the electron turns at $a_j$ and runs to $x_{j0}$, the mcf should
   be assumed to be given by $W_j(x_j) - 2W_j(a_j) + 2W_j(b_j)$.
  The wave function for $x_j (\le x_{j0})$ is written as
 \begin{eqnarray}
     - \Phi_j^*(x_j)e^{ i 2W_j(b_j) }
     + \Phi_j(x_j) e^{ i \{ - 2W_j(a_j) + 2W_j(b_j) \} }.
                            \label{eq:xjaj}
\end{eqnarray}
 This is finite, zero, at $a_j$.
 The wave functions~\eqref{eq:xjbj} and ~\eqref{eq:xjaj} are finite
 for $a_j \le x_j \le b_j$ for any `mode' with real values
  of parameters.
 They are not always equal at $x_{j0}$.
As a wave function, it might be multi-valued.
If and only if $2[W_j(b_j) - W_j(a_j)]$ is a multiple of $2 \pi$,
 they are equal and constitute a stationay wave function
 in $a_j \le x_j \le b_j$.

In what follows, the wave equation~\eqref{eq:Schroe} is analized
 in the polar coordinate system, $(r, \theta, \phi)$.
The Hamilton characteristic function is summarized as follows:
\begin{eqnarray}
 W_{\rm cl}(r, \theta, \phi, E, l, \mu) = W_{r, \rm cl}(r, E, l)
  + W_{\theta, \rm cl}(\theta, l, \mu) + \mu \hbar,
\end{eqnarray}
where $W_{r, \rm cl}$ and $W_{\theta, \rm cl}$ are determined from
equations
\begin{eqnarray}
  \frac{1}{2m} \left[
   \left(\frac{\partial W_{r, \rm cl}}{\partial r} \right)^2
  + \frac{l^2 \hbar^2}{r^2} \right] - \frac{e^2}{r} = E,  \\
   \left(\frac{\partial W_{\theta, \rm cl}}{\partial \theta} \right)^2
  + \frac{\mu ^2 \hbar ^2}{\sin ^2 \theta} = l^2 \hbar ^2.
\end{eqnarray}
Here, $E$ stands for the energy and will be expressed in terms of
 $\eta_{s}$,~\eqref{eq:eta_s}, for the scattering state or
 $\eta$,~\eqref{eq:eta}, for the bound state.
 Parameters $l$ and $\mu$ are real numbers and $l \hbar $ stands for
 the angular momentum and $\mu \hbar $ is one of its components.
 These expressions have been introduced for convenience for comparison
 with the quantum theory.

\section{ Scattering state} \label{sec:SCAT}
The scattering state of an electron in the Coulomb potential is analyzed
 in the spherical polar coordinate system.
 The wave function $\Psi ({\bf r},t)$ is expressed in the spherical
 polar coordinates with mode parameters,
  $E, \nu$  and $\mu $ as
\begin{eqnarray}
  \Psi({\bf r},t) &=& \exp \left( -i E t / \hbar \right)
   \Phi ({\bf r},E), \label{eq:Psirt} \\
  \Phi ({\bf r},E) &=& R(r,E,\nu )Y(\theta ,\nu , \mu )
                     \exp(i\mu \phi ).  \label{eq:PhirEnm}
\end{eqnarray}
Constant $E$ stands for the energy and $\hbar \nu$  for the orbital
 angular momentum, and $\hbar \mu$  represents the component of
 the angular momentum along the polar axis.
When $\nu$ and $\mu$ are integral numbers, they are usual azimuthal and
 magnetic quantum numbers as will be seen in Eqs.~\eqref{eq:Ytheta}
and ~\eqref{eq:urEnu}.~\cite{SchiffS4}

The mcf expressed in terms of the spherical polar coordinates are obtained
 as follows.
The function $Y(\theta ,\nu , \mu )$ satisfies the differential equation
\begin{eqnarray}
\left[ {d^2 \over d\theta ^2} + \cot \theta {d \over d\theta }
 + \nu (\nu +1) - {\mu ^2 \over \sin^2\theta } \right]
  Y(\theta ,\nu , \mu ) = 0.
                                   \label{eq:Ytheta}
\end{eqnarray}
The solution is a linear combination of linearly independent
 associated Legendre functions,
 $ P_{\nu}^{\mu} (\cos \theta )$ and
  $ Q_{\nu}^{\mu} (\cos \theta )$.~\cite{Erdely}
A traveling wave in the $\theta$ coordinate space is given by
\begin{eqnarray}
 Q_{\nu}^{\mu} (\cos \theta ) + i\frac{\pi}{2}P_{\nu}^{\mu} (\cos \theta )
 &=&  \left|  \frac{2}{\pi } \sin \theta
       \frac{\partial W_{\theta}}{\partial \theta} \right|^{-1/2}
     \exp \left [ i W_{\theta}(\theta ,\nu , \mu ) \right ]
					   \nonumber\\
  &\equiv&  \sqrt{P_{\theta}} e^{i W_{\theta}}.
                                     \label{eq:Yteta}
\end{eqnarray}

The mcf for the $\theta$ component should be determined as
\begin{eqnarray}
 W_{\theta} (\theta ,\nu ,\mu ) = \arctan \left[ {\pi \over 2}
  {P_{\nu}^{\mu} (\cos \theta ) \over Q_{\nu}^{\mu} (\cos \theta )}
   \right],                 \label{eq:Wtheta}
\end{eqnarray}
because of the similarity to the characteristic function
$W_{\theta, \rm cl}$ in the classical region and the validity of
the results derived from this as will be seen in the following.

An expression for function $\tan(W_{\theta} )$
 is written for $0 < \theta < \pi$ as~\cite{AbSt8}
\begin{eqnarray}
 \lefteqn{
 {\pi P_{\nu}^{\mu} (\cos \theta ) \over 2Q_{\nu}^{\mu} (\cos \theta )} }
                                                    \nonumber \\
  &=& \sum_{k = 0}^{\infty }{(1/2+\mu )_k (1+\nu +\mu )_k \over
  k! (\nu +3/2)_k} \sin [ (2k+\nu +\mu +1)\theta ] \Bigg/
                                                    \nonumber \\
  & & \sum_{k = 0}^{\infty }{(1/2+\mu )_k (1+\nu +\mu )_k \over
  k! (\nu +3/2)_k} \cos [ (2k+\nu +\mu +1)\theta ].
\end{eqnarray}
 The value of $W_{\theta}$ at $\theta = 0$ or $ \pi$
is~\cite{Erdely}
\begin{eqnarray}
 W_{\theta} (0) \equiv W_{\theta}(0, \nu, \mu) = \pi \mu, \quad
  W_{\theta}(\pi) \equiv W_{\theta}(\pi, \nu, \mu) = \pi \nu.
			\label{eq:Wtheta0pi}
\end{eqnarray}
By the asymptotic expansion of the Legendre functions for
 $\nu \gg 1$,~\cite{AbSt8} it can be obtained that
 \begin{eqnarray}
  W_{\theta}(\theta, \nu, \mu) \approx
    \left( \nu + \FRAC{1}{2} \right) \theta
     + \left( \FRAC{1}{4} + \FRAC{1}{2}\mu \right) \pi,
 \mbox{\hspace*{.5cm} ($\epsilon < \theta < \pi - \epsilon,\;
   \epsilon > 0$)}.
			\label{eq:Wthetaprx}
\end{eqnarray}

 Both the derivative $\partial W_{\theta} /\partial \nu $
 and $\partial W_{\theta} /\partial \mu $  are monotonic as a function
 of $\theta $ and
  very similar to those of the characteristic function
   $W_{\theta, \rm cl}$ in a restricted (or classical) region
   as found  by a computer calculation.

 Radial wave function satisfies the differential equation
\begin{eqnarray}
 \left[ {d^2 \over dr^2} - {\nu (\nu +1) \over r^2} + {2m \over \hbar ^2}
 \left( {e^2 \over r} + E \right) \right] u(r) = 0,
                              \label{eq:urEnu}
\end{eqnarray}
where $u(r,E,\nu ) = r R(r,E,\nu )$.

 With $E$ positive the linearly independent solutions are
\begin{eqnarray}
 u_M &=& e^{-i\rho} \rho^{\nu +1} M(\nu +1+i\eta_s ,2\nu +2, i2\rho),
                                     \label{eq:uM} \\
 u_V &=& e^{-i\rho} \rho ^{\nu +1} V(\nu +1+i\eta_s ,2\nu +2, i2\rho),
\end{eqnarray}
where
\begin{eqnarray}
 \rho = \sqrt{2mE \over \hbar ^2}r, \quad
 \eta_s = \frac{e^2}{\hbar}\sqrt{\frac{m}{2 E}}.
 			\label{eq:eta_s}
 \end{eqnarray}
 Function $V(a,b,z)$ is defined for convenience~\cite{JOSAA95}
\begin{eqnarray}
 V(a,b,z)
  &=& \Gamma(a) \left[ U(a, b, z)
   - \cos \pi a \frac{\Gamma(b-a)}{\Gamma(b)} M(a, b, z) \right]
	   					\nonumber\\
  &=& -\pi \cot \pi b {\Gamma(b-a) \over
                       \Gamma(b) \Gamma(1-a)} M(a,b,z) \nonumber \\
  & &  + \Gamma(b-1) z^{1-b} M(1+a-b,2-b,z).   \label{eq:defV}
\end{eqnarray}
 Functions $M(a,b,z)$  and $U(a,b,z)$ are the Kummer
  functions.~\cite{AbraSteg}

 For the far region from the center of the potential, $\rho  \gg 1 $,
 by putting $a = \nu  + 1 + i\eta_s $ and $b = 2\nu  + 2,$
  it holds ~\cite{AbraSteg,Slater} that
\begin{eqnarray}
 M(a, b, i2\rho) &\simeq&  e^{i\rho} \Gamma (b) \left[
 {e^{-i(\rho - \pi a/2)} \over \Gamma (a^*)}(2\rho )^{-a}
   \left(1 + \frac{i a(1 - a^*)}{2 \rho} \right)
                                             + c.c. \right], \\
V(a,b,i2\rho ) &\simeq&  e^{i\rho} \left[ -i \sin (\pi a) G(\rho ,a)
 - \cos (\pi a) G(\rho ,a)^{\ast} \right],
\end{eqnarray}
where
\begin{eqnarray}
 G(\rho ,a) = \Gamma (a)e^{-i(\rho - \pi a/2)}(2\rho )^{-a}
      \left(1 + \frac{i a(1 - a^*)}{2 \rho} \right),
\end{eqnarray}
and $G(\rho ,a)^*$ is the complex conjugate (c.c.) of $G(\rho ,a)$.
These asymptotic forms indicate that  the linear
combination of functions $M$ and $V$ producing an outgoing traveling wave
in the far region from the origin should be written as
\begin{eqnarray}
u = e^{-i\rho} \rho^{\nu +1} \left[ V(a, b, i2\rho)
  + iM(a, b, i2\rho) \sin (\pi a)
  {\Gamma (a)\Gamma (b-a) \over \Gamma (b)} \right].
\end{eqnarray}
By equations mentioned above, this leads to
\begin{eqnarray}
 u &\approx& e^{- i\pi \nu} e^{\FRAC{1}{2}\pi \eta_s} 2^{-1-\nu}
   |\Gamma(a)| \exp[i(\rho + \eta_s \log (2\rho) - \arg \Gamma(a)
    - \FRAC{1}{2}\pi (1 + \nu)) ] 		\nonumber\\
 & & \times \left(
   1 + i \frac{\nu(1 + \nu) + \eta_s^2}{2\rho} - \frac{\eta_s}{2\rho}
     + O(\rho^{-2}) \right),
\end{eqnarray}
for $\rho $ large.~\cite{GORDON,MottMass}
 It would suggest that the traveling wave in the $r$ coordinate space
 should be given by
\begin{eqnarray}
 u &=& u_V + iu_M \sin (\pi a) {\Gamma (a)\Gamma (b-a) \over \Gamma (b)}
  \equiv    e^{-i\pi \nu } (u_R + iu_I) \nonumber \\
 &=&  e^{-i\pi \nu} \sqrt{u_R^2 + u_I^2}
  \exp \left( i\arctan {u_I \over u_R} \right)
  \equiv e^{-i\pi \nu} \rho \sqrt{P_r} e^{i W_r},
				   \label{eq:wfmotion}
\end{eqnarray}
and thus the mcf in the $r$ coordinate is given by
\begin{eqnarray}
 W_r(r,E,\nu ) = \arctan {u_I \over u_R}.
                   \label{eq:Wr}
\end{eqnarray}
Here, functions $u_R$ and $u_I$ have been assumed to be real.
 The parameter $\nu$ is a real number.
The ratio $ u_R / u_I $ is written as
\begin{eqnarray}
 \frac{u_I}{u_R} &=&
  \frac{\cos (\pi b) - \exp(-2\pi \eta_s)}{\sin (\pi b)} - H,
			                    \label{eq:uRuI}    \\
 H &=& 2^{2-b} {\Gamma (b) \Gamma (b-1) \over |\Gamma (a)|^2}
      \exp(-\pi \eta_s) \rho ^{1-b} {M(a+1-b, 2-b, i2\rho) \over
      M(a, b, i2\rho)}
\end{eqnarray}
Function $H$ is real since it holds  by Kummer's
 transformation~\cite{AbraSteg} that
\begin{eqnarray}
 & &{} e^{-i\,\rho} M(a, b, i2\rho) = e^{i\,\rho} M(a, b, i2\rho)^*, \\
 & &{}  e^{-i\,\rho} M(a+1-b, 2-b, i2\rho) =
     e^{i\,\rho} M(a+1-b, 2-b, i2\rho)^*.
\end{eqnarray}

Function~\eqref{eq:PhirEnm} with expressions~\eqref{eq:Yteta}
 and~\eqref{eq:wfmotion} specifies a traveling wave associated
 with the motion of an electron
 in the `mode' $(E, \nu, \mu)$ in the scattering state.

In the far region from the origin the mcf is approximated  as
\begin{eqnarray}
  W_r(r, E, \nu )
   &\simeq&  \rho  + \eta_s \log 2\rho
      - \arg \Gamma(a) - {}^1\!/\!_2 \pi(1 + \nu)
       + \frac{\nu(1 + \nu) + \eta_s^2}{2 \rho}  \nonumber\\
  & &{} + O( \rho^{-2}).      \label{eq:Wrfar}
\end{eqnarray}
 This is nearly equal to the corresponding Hamilton characteristic
  function.~\cite{GORDON,MottMass}
  It indicates the validity of the definition of the mcf~\eqref{eq:Wr}.
For $r$ small, it is obtained that
\begin{eqnarray}
   W_r(0) \equiv W_r(0, E, \nu) =  \FRAC{1}{2} \pi.
\end{eqnarray}

 The  mcf in the spherical polar coordinates is summarized  as
\begin{eqnarray}
 W(r,\theta ,\phi ,E,\nu ,\mu ) &=& \pm W_r(r,E,\nu )
     \pm W_{\theta} (\theta ,\nu ,\mu ) \pm  \mu \phi, \label{eq:mcfSch}
\end{eqnarray}
where the sign should be adopted according to the direction of the
motion of the particle in each coordinate.
 Equations of motion~\eqref{eq:MotionS} with $ r, \theta, \phi, \nu$
 or $\mu$ in place of $x_1, x_2, x_3, \alpha$ or $\beta$,
 respectively, lead to a  trajectory of the electron.

To get the trajectory, it is necessary  to calculate derivatives
$\partial W_r / \partial E $ and $\partial W_r /   \partial \nu $.
 It is obtained from Eqs.~\eqref{eq:Wr}
 and (\ref{eq:uRuI}) that
\begin{eqnarray}
 \frac{\partial W_r}{\partial \nu } &=&  {1 \over 1 + (u_I/u_R)^2}
 \frac{\partial}{\partial \nu } \left( \frac{u_I}{u_R} \right),
					 \nonumber \\
  {\partial \over \partial \nu }\left( {u_I \over u_R} \right)
  &=& 2\pi \frac{\exp(-2\pi \eta_s) \cos \pi b - 1}{ \sin ^2\pi b}
   - H \Bigg[ 2 \left( \psi(b-1) + \psi(b) - \log 2\rho \right)
    				 \nonumber \\
   & &{}- \psi(a)- \psi(a^*) - {1 \over M_2} \left(
    \frac{\partial M_2}{\partial a}
   + 2 \frac{\partial M_2 }{\partial b} \right)		\nonumber\\
   & &{} - {1 \over M} \left( \frac{\partial M }{ \partial a}
    + 2 \frac{\partial M }{\partial b} \right) \Bigg],
\end{eqnarray}
where $M = M(a, b, i2\rho ), \psi(b)$ is the psi (digamma) function and
\begin{eqnarray}
 M_2 = M(a+1-b, 2-b, i2\rho ) = M(-\nu + i\, \eta_s, -2\nu, i2\rho ).
\end{eqnarray}

For example, a trajectory of an electron incident from a point
 distant from the origin of the potential,
 ($\rho_0, \theta_0, \phi_0)$, and scattered to another distant
 point is considered.
To be specific, that $\rho_0 = \infty$ and $\theta_0 =0$ is
 assumed.
For the trajectory from $\rho_0$ to the origin, the mcf is written as
\begin{eqnarray}
  W(r, \theta, \phi, E, \nu, \mu) = - W_r + W_{\theta}
		  + W_{\phi},
		  \label{eq:mcfsctin}
\end{eqnarray}
where $W_{\phi} = \mu \phi$.
The trajectory is given by the equations (\ref{eq:MotionS}), or
\begin{eqnarray}
 \frac{\partial}{\partial \nu} \left( - W_r + W_{\theta} \right)
 &=& \frac{\partial}{\partial \nu} \left( - W_r(\infty) + W_{\theta}(0)
  \right) = \frac{\partial}{\partial \nu} \arg \Gamma(a)
  + \FRAC{1}{2} \pi,			\\
 \frac{\partial}{\partial \mu} \left( W_{\theta} + W_{\phi} \right)
 &=& \frac{\partial}{\partial \mu} W_{\theta}(0) + \phi_0
   = \pi + \phi_0.	\label{eq:phitheta}
\end{eqnarray}
For the path from the origin to $\rho_{\pi} \equiv \rho(\theta=\pi)$,
the mcf is given by
\begin{eqnarray}
  W(r, \theta, \phi, E, \nu, \mu) =  W_r + W_{\theta}
			   + W_{\phi} - 2 W_r(0).
			\label{eq:r0torpi}
\end{eqnarray}
This has been taken to be continuous to the incident mcf,
 (\ref{eq:mcfsctin}), at the origin.

The trajectory equation is written as
\begin{eqnarray}
\lefteqn{ \frac{\partial}{\partial \nu} \left(
 W_r + W_{\theta} - 2 W_r(0) \right) } 		 \nonumber\\
 &=& \frac{\partial}{\partial \nu} \arg \Gamma(a)
   + \FRAC{1}{2} \pi,
\end{eqnarray}
 and Eq.~(\ref{eq:phitheta}).
Since function $\partial_{\nu} W_r$ shows monotonic decrease
with respect to $\rho$ while $\partial_{\nu} W_{\theta}$ does
 monotonic increase with respect to $\theta$
 as proved by the computer calculation,
there is a point $\rho = \rho_{\pi}$ where $\theta$ takes $\pi$.

For the trajectory from $\rho_{\pi}$ to a distant scattered point,
 the mcf is written as
\begin{eqnarray}
 W(r, \theta, \phi, E, \nu, \mu) =  W_r - W_{\theta}
		   + W_{\phi} - 2 W_r(0) + 2 W_{\theta}(\pi).
\end{eqnarray}
The last term has been added to for the continuity of $W$ at
the point $\rho = \rho_{\pi}$ to the expression (\ref{eq:r0torpi}).

The trajectory is given by
\begin{eqnarray}
 \frac{\partial}{\partial \nu} \left( W_r - W_{\theta} - 2 W_r(0)
    + 2 W_{\theta}(\pi) \right)
 &=& \frac{\partial}{\partial \nu} \arg \Gamma(a)
      + \FRAC{1}{2} \pi,			\label{eq:scattered} \\
 \phi &=& \frac{\partial}{\partial \mu}W_{\theta} + \pi + \phi_0.
\end{eqnarray}

\begin{figure}[htbp]
\begin{center}
  \includegraphics[width=9.5cm]{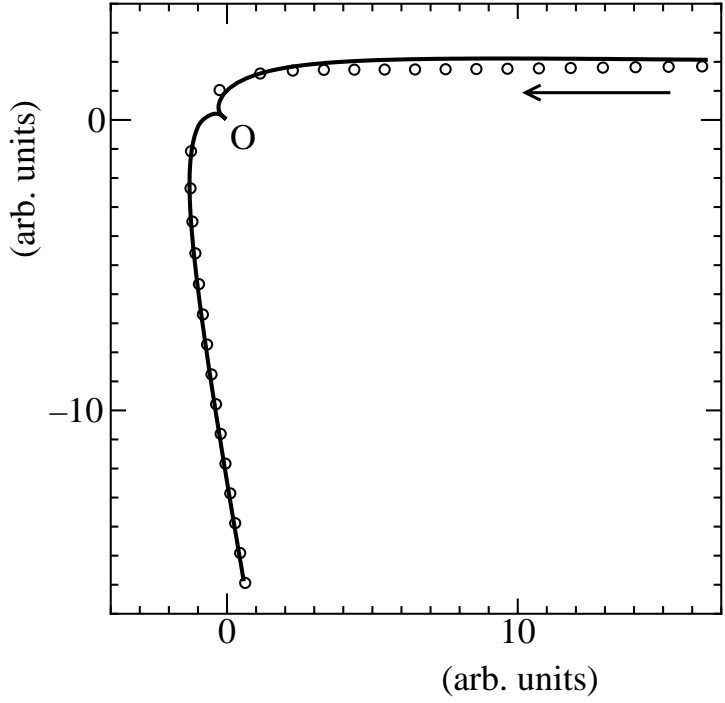}
\end{center}
 \caption{\small  A trajectory of an electron of the hydrogen atom
  in the scattering `mode' with parameters $\eta_s  = 3.3, \nu  = 2.2,
 \mu  = 1.1,$ (solid line) and the classical orbit with 
 $\eta_s  = 3.3, l = 2.7, \mu  = 1.1,$ (circle).
}
\label{fig:Coulsc}
\end{figure}

The equations of the trajectory prescribed above are more easily
 seen if the approximate relation (\ref{eq:Wthetaprx}) is used.
It shows that coordinate $\phi$ is constant except the point
of $\rho_{\pi}$ where  $\theta=\pi$.
This indicates that the trajectory is almost on a plane like the classical
 orbit.

An example of a trajectory of the electron with parameters
 $\eta_s  = 3.3, \nu  = 2.2, \mu  = 1.1,$, projected on a plane, is
 shown with the corresponding classical orbit with
 $\eta_s  = 3.3, l = 2.7, \mu  = 1.1,$
  in Fig.~\ref{fig:Coulsc}.

  The corresponding classical orbit, as is well known, is very similar
   to this trajectory except the neighborhood of the origin.

\section{Cross section}\label{sec:crossect}
Along the line of thought of classical mechanics, the cross section
will be detrermined by the mode trajectory.

The differential cross section for the classical orbits of uniform
incident beam is given by~\cite{Goldstein}
\begin{eqnarray}
 \sigma(\theta)_{\rm cl}
  = \frac{e^4}{16 E^2} \frac{1}{\sin^4 (\theta /2)}
 = \left(\frac{\hbar^2}{me^2} \right)^2 \eta_s^2 \frac{\eta_s^2}{4}
   \frac{1}{\sin^4 (\theta /2)},
   				\label{eq:dfcrosec_cl}
\end{eqnarray}
where $\theta$ is the scattering angle of the electron beam
 and parameter $\eta_s $ (\ref{eq:eta_s}) has been used.
 This is equal to that obtained in quantum mechanics not to mention.
The scattering angle is expressed, in terms of the angular momentum
 $l \hbar$ and energy $E$, as
\begin{eqnarray}
  \theta = 2 \arctan \left(
       \sqrt{\frac{m}{2E}} \frac{e^2}{l \hbar} \right).
\end{eqnarray}
The impact parameter, $s$, is related to the angular momentum as
 $s = l \hbar /\sqrt{2 m E}$.

Equation~(\ref{eq:scattered}) leads to the scattering (or deviated)
 angle of the electron trajectory, by taking $\rho = \infty$ and
 using the approximate relation~(\ref{eq:Wthetaprx}), or
 $\partial_{\nu} W_{\theta} \approx \theta$,
\begin{eqnarray}
 \theta_{\rm sc} = \pi
  - \frac{\partial}{\partial \nu} W_{\theta, \rho = \infty}
  = 2 \frac{\partial}{\partial \nu} \arg \Gamma(a).
		\label{eq:scatangl}
\end{eqnarray}
In the remote region from the origin, equation (\ref{eq:Wrfar}) shows that
 the difference between two positions along the trajectory satisfies
\begin{eqnarray}
  \frac{d \theta}{d \rho} = \frac{\partial}{\partial \rho}
  \frac{\partial}{\partial \nu} W_r = - \frac{\nu + \FRAC{1}{2}}{\rho^2}.
\end{eqnarray}
Integration gives rise to $\rho \theta = \nu + \FRAC{1}{2}$.
It is thus obtained that the impact parameter of the trajectory
 is given by
\begin{eqnarray}
  s = r \theta = (\nu + \FRAC{1}{2})\hbar / \sqrt{2 m E}.
\end{eqnarray}
This indicates that $(\nu + \FRAC{1}{2}) \hbar$ corresponds to
 $ l \hbar$, angular momentum in the sense of classical mechanics,
 and $\nu$ should be greater than $- \FRAC{1}{2}$.

\begin{figure}[htbp]
\begin{center}
  \includegraphics[width=7.8cm]{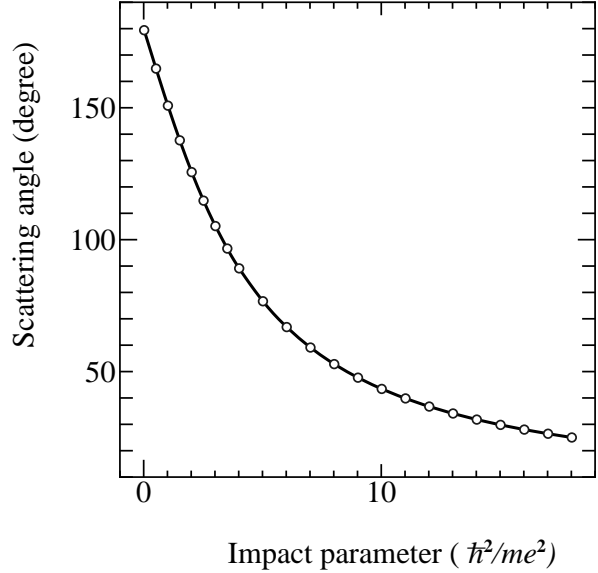}
\end{center}
\caption{ Scattering angle as a function of the impact parameter
  of an incident beam for the classical orbit (circle) and
 for the trajectory (solid line) with $\eta_s = 2$.}
\label{fig:Scatangl}
\end{figure}

An example of the scattering angle of an incident beam as a function
 of the impact parameter for the classical orbit or for the trajectory
 is shown in Fig.~\ref{fig:Scatangl}.

The differential cross section for the trajectories of uniform
 incident beam of the electron is obtained in the same way
  as the classical one,
by using (\ref{eq:scatangl}),
\begin{eqnarray}
 \sigma(\theta_{\rm sc}) &=&
  \frac{s}{\sin \theta_{\rm sc}}
     \left | \frac{d s}{d \theta_{\rm sc}} \right |
   = \frac{(\nu + \FRAC{1}{2})\hbar^2}{2 m E}
       \frac{1}{\sin \theta_{\rm sc}}
         \left | \frac{d \nu}{d \theta_{\rm sc}} \right |
  			\nonumber\\
  &=&  \left(\frac{\hbar^2}{me^2} \right)^2 \eta_s^2
   \frac{(\nu + \FRAC{1}{2})}{2 \sin \theta_{\rm sc}}
   \left | \frac{\partial^2}{\partial \nu^2} \arg \Gamma(a)
               \right |^{-1}.
       			\label{eq:dfcrosec_tr}
\end{eqnarray}
Parameter $\nu$ in the right hand side should be expressed
in terms of $\theta_{\rm sc}$ through (\ref{eq:scatangl}).

\begin{figure}[htbp]
\begin{center}
  \includegraphics[width=7.8cm]{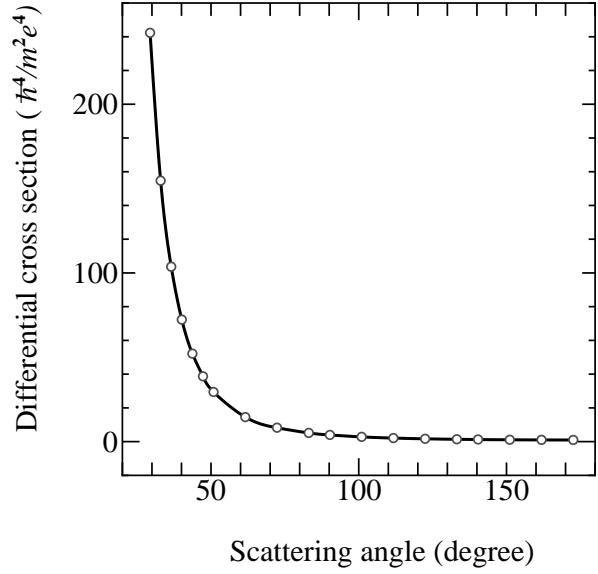}
\end{center}
\caption{ Differential cross section as a function of the scattering
 angle  of an incident beam for the classical orbit (circle) and
 for the trajectory (solid line) with $\eta_s = 2$.}
\label{fig:Difcrosect}
\end{figure}

An example of the differential cross section of an electron beam
  for the classical orbit and that for the trajectory are shown
  in Fig.~\ref{fig:Difcrosect}.

The figures indicate that the similarity between the cross sections
 of the classical orbits and the trajectories  is, as it were,
  complete.

\section{Flux}\label{sec:flux}
The usual wave function for the Coulomb scattering is given, by
Gordon, as~\cite{GORDON,MottMass}
\begin{eqnarray}
\psi(r, \theta) = \sum_{l=0}^{\infty} (2 l + 1) i^l
 \exp(- i \arg \Gamma(1 + l + i \eta_s)) L_l(r) P_l (\cos \theta)
\end{eqnarray}
where
\begin{eqnarray}
 L_l(r) = \frac{2^{l+1} \exp( - \pi/2 \eta_s)}{
  \rho |\Gamma(1 + l + i \eta_s)|}u_R(\rho).
\end{eqnarray}
By using (\ref{eq:Yteta}) and (\ref{eq:wfmotion}), it can be composed of a linear combination
 of the travelling waves with parameters $\eta_s, \nu = l$ and
 $\mu = 0$
\begin{eqnarray}
\psi(r, \theta) &=& \sum_{l=0}^{\infty} (2 l + 1) \frac{(2i)^l
 \exp(- \FRAC{1}{2} \pi \eta_s)}{ \pi \Gamma(1 + l + i \eta_s)}
  \left( \sqrt{P_r}e^{- iW_r} - \sqrt{P_r} e^{i[W_r - 2 W_r(0)]} \right)
				\nonumber\\
 & &{} \times
  \left(  \sqrt{P_{\theta}} e^{iW_{\theta}}
       - \sqrt{P_{\theta}} e^{-i[W_{\theta} - 2 W_{\theta}(\pi)]} \right).
\end{eqnarray}
For $r$ large, it is approximately expressed as~\cite{GORDON,MottMass}
\begin{eqnarray}
\psi(r, \theta) &\simeq&
 \exp[i (\rho \cos \theta - \eta_s \log \rho (1 - \cos \theta)]
 + \frac{\eta_s}{\rho (1 - \cos \theta)}	\nonumber\\
 & &{}\times
 \exp \left( i [\rho + \eta_s \log \rho (1 - \cos \theta)
	+ \pi - 2 \arg \Gamma(1 + i \eta_s)] \right).
		\label{eq:psiGappr}
\end{eqnarray}
The flux (or current) of beam of particles is defined by
\begin{eqnarray}
   {\bf j}
  = \frac{1}{ 2 m } \psi^* \frac{ \hbar}{i}\nabla \psi + c.c.
\end{eqnarray}
The first term of the right-hand side of the wave
function (\ref{eq:psiGappr}) gives rise
 to the incident flux propagating along the $z$ axis for $r$ large
\begin{eqnarray}
 {\bf j}_{\rm in} = \frac{\hbar k}{m} \hat{z}.
\end{eqnarray}
The second term, on the other hand, gives rise to the outgoing flux
\begin{eqnarray}
 {\bf j}_{\rm out} = \frac{\hbar k}{m}
 \frac{\eta_s^2}{\rho^2 (1 - \cos \theta)^2} \hat{\bf r}.
\end{eqnarray}
The number of particles in the outgoing flux in a solid angle is
\begin{eqnarray}
 \frac{ r^2 {\bf j}_{\rm out} \cdot \hat{\bf r}}{\hbar k / m}
 = \left( \frac{\hbar^2 \eta_s}{m e^2} \right)^2
   \frac{\eta_s^2}{4 \sin^4 \FRAC{1}{2} \theta},
\end{eqnarray}
which gives the differential cross section.

According to the interpretation envisaged with Gordon's work,
the deviated beam should be obtained through the scattering
of incident particles with various impact parameters.
The outgoing flux is considered to be an assemblage of particles which,
 in classical mechanicsl sense, should be governed
 by a plausible causal dynamics.

In the present treatment, the associated wave with the motion of an
electron with $\eta_s, \nu$ and $\mu = 0$, may be expressed, for $r$ large,
 as
\begin{eqnarray}
\psi(r, \theta)_{\nu} &\simeq&
       \sqrt{P_r P_{\theta}} e^{- i (W_r - W_{\theta})}
   	+ \sqrt{P_r P_{\theta}} e^{i [W_r - W_{\theta} + (2\nu - 1)\pi]}
   					\nonumber\\
  & &{} \equiv \psi_{\nu, \rm in} + \psi_{\nu, \rm out}.
  			\label{eq:nu_wave}
\end{eqnarray}
It holds approximately that
\begin{eqnarray}
 P_r \simeq \rho^{-2} 2^{-2\nu + 2} e^{\pi \eta_s}
  |\Gamma(1 + \nu + i \eta_s)|^2, \quad
 P_{\theta} \simeq \frac{\pi}{(2\nu + 1) \sin \theta}.
\end{eqnarray}

The wave for a beam of particles with various impact marameters
should be written as
\begin{eqnarray}
 \psi(r, \theta) = \sum_{\nu} a_{\nu} \psi(r, \theta)_{\nu},
\end{eqnarray}
where $a_{\nu}$ is an arbitrary constant which will indicate
the intensity and initial phase of the wave (\ref{eq:nu_wave}).

The flux is written
\begin{eqnarray}
 {\bf j}  = \frac{1}{ 2m} \sum_{\nu} \sum_{\nu'}
   a_{\nu}^* a_{\nu'} \psi^*_{\nu} \frac{\hbar}{i} \nabla \psi_{\nu'}
     + c.c.
  \equiv  \sum_{\nu} \sum_{\nu'} {\bf j}_{\nu, \nu'}.
\end{eqnarray}
For $r$ large, the incident waves $\psi_{\nu, \rm in}$'s give
rise to the incident flux
\begin{eqnarray}
 {\bf j}_{\nu \nu', \rm in} &\simeq& \hat{\bf r} \frac{\hbar k}{m}
 \frac{|a_{\nu} a_{\nu'}|}{\rho^2 \sin \theta_{\rm in}}
 \frac{\pi \exp(\pi \eta_s)}{\sqrt{(2\nu + 1)(2\nu' + 1)}}
 \frac{|\Gamma(1 + \nu + i \eta_s) \Gamma(1 + \nu' + i \eta_s)|}{2^{\nu + \nu' + 2}}	\nonumber\\
& &{}\times \cos \bigl( -\arg \Gamma(1 + \nu + i \eta_s) + \arg \Gamma(1 + \nu'      + i \eta_s)					\nonumber\\
& &{}\quad  - (\theta + \FRAC{1}{2}\pi )(\nu - \nu')
		 + \arg(a_{\nu}^* a_{\nu'}) \bigr).
\end{eqnarray}
Similarly, the outgoing waves  $\psi_{\nu, \rm out}$'s give
rise to the scattered flux
\begin{eqnarray}
 {\bf j}_{\nu \nu', \rm out} &\simeq& \hat{\bf r} \frac{\hbar k}{m}
 \frac{|a_{\nu} a_{\nu'}|}{\rho^2 \sin \theta_{\rm out}}
 \frac{\pi \exp(\pi \eta_s)}{\sqrt{(2\nu + 1)(2\nu' + 1)}}
 \frac{|\Gamma(1 + \nu + i \eta_s) \Gamma(1 + \nu' + i \eta_s)|}{2^{\nu + \nu' + 2}}	\nonumber\\
& &{}\times \cos \bigl( \arg \Gamma(1 + \nu + i \eta_s) - \arg \Gamma(1 + \nu' + i \eta_s)					\nonumber\\
& &{}\quad
 + (\theta + \FRAC{1}{2}\pi )(\nu - \nu') - 2\pi (\nu - \nu')
 + \arg(a_{\nu}^* a_{\nu'}) \bigr).
\end{eqnarray}
For $\nu \simeq \nu'$,
\[  \arg \Gamma(1 + \nu + i \eta_s) - \arg \Gamma(1 + \nu' + i \eta_s)
   \simeq
 (\nu - \nu') \partial_{\nu} \arg \Gamma(1 + \nu + i \eta_s).	\]
From (\ref{eq:scatangl}),
\[   \theta_{\rm out} = \pi - (\theta_{\rm scat} + \theta_{\rm in})
     = \pi - 2 \partial_{\nu} \arg \Gamma(1 + \nu + i \eta_s)
   - \theta_{\rm in},	\]
where $\theta_{\rm in}$ is written as $\theta_0$ in a preceding section.
Thus, the argument of cosine in ${\bf j}_{\nu \nu', \rm out}$ with
 $\theta_{\rm out}$ equals that of cosine in  ${\bf j}_{\nu \nu', \rm in}$
 with $\theta_{\rm in}$,
\begin{eqnarray}
 \lefteqn{ \arg \Gamma(1 + \nu + i \eta_s) - \arg \Gamma(1 + \nu' + i \eta_s)
 + (\theta + \FRAC{1}{2}\pi )(\nu - \nu') }			\nonumber\\
 & &{}  - 2\pi (\nu - \nu') + \arg(a_{\nu}^* a_{\nu'}) 		\nonumber\\
 & &{} \simeq (\nu - \nu') \left[
  \partial_{\nu} \arg \Gamma(1 + \nu + i \eta_s)
 + \theta_{\rm out} - \FRAC{3}{2} \pi \right] +  \arg(a_{\nu}^* a_{\nu'})
							\nonumber\\
 & &{}= - (\nu - \nu') \left[ \partial_{\nu} \arg \Gamma(1 + \nu + i \eta_s)
 + \theta_{\rm in} - \FRAC{1}{2} \pi \right] +  \arg(a_{\nu}^* a_{\nu'}).
\end{eqnarray}
It is seen thus that for $\rho \gg 1$
\begin{eqnarray}
 {\bf j}_{\nu \nu', \rm in} \cdot (- \hat{\bf r })\,
   \rho^2 \sin \theta_{\rm in}
 = {\bf j}_{\nu \nu', \rm out} \cdot \hat{\bf r}\,
  \rho^2 \sin \theta_{\rm out}.
		\label{eq:fluxconsv}
\end{eqnarray}
If it is  assumed that the incident beam  ${\bf j}_{\nu \nu'}$
 with $\nu'$ clearly different from $\nu$ is
incoherent, $< \cos \arg (a_{\nu}^* a_{\nu'}) > \mbox{  (average over
 the initial phases)} = 0$, the total flux may be expressed as
\begin{eqnarray}
       \sum_{\nu, \nu'} {\bf j}_{\nu, \nu'}
   = \sum_{\nu \simeq \nu'} {\bf j}_{\nu, \nu'}.
\end{eqnarray}
The current $ {\bf j}_{\nu, \nu'}.$ with $ \nu' \simeq \nu$ could be
discriminated from the one $ {\bf j}_{\mu, \mu'}$ with $ \mu' \simeq \mu$
 clearly different from $\nu$ because of the difference of the impact
 parameter or the angular momentum.
Then, expression (\ref{eq:fluxconsv}) means that the flux density of particles
along the trajectory is conserved for $r$ large, or  the flux density of
the incident beam is equal to that of the outgoing beam.
Therefore, the cross section for the incoherent incident beam can
be calculated like the classical mechanical beam
 as (\ref{eq:dfcrosec_tr}).

\section{Bound state }\label{sec:HYDRSCH}
 An electron trajectory in a hydrogen atom in a bound state is analyzed.
  The solution of the form
  of expressions (\ref{eq:Psirt}) and (\ref{eq:PhirEnm})
 with  energy $E$ negative is sought.
 The mcf in the $\theta$ coordinate space, $W_{\theta}$, is the same in
 the preceding section, (\ref{eq:Wtheta}).

 The traveling wave in the radial coordinate is obtained
  from Eq.~\eqref{eq:wfmotion}
 with replacement of $i \eta_s$  or $i \rho$  by $-\eta$ or $\rho $,
 respectively.
 By writing $u(r,E,\nu ) = r R(r,E,\nu )$,  a traveling wave is obtained
\begin{eqnarray}
  u = u_V + iu_M \sin \pi a {\Gamma (a)\Gamma (b-a) \over
    \Gamma (b)},  \label{eq:NRVM}
\end{eqnarray}
where $a = \nu  + 1 - \eta ,$ and $b = 2\nu  + 2.$
Functions $u_M$ and $u_V$ are given by
\begin{eqnarray}
 u_M &=& e^{-\rho} \rho ^{\nu +1} M(\nu +1-\eta ,2\nu +2,2\rho ), \\
 u_V &=& e^{-\rho} \rho ^{\nu +1} V(\nu +1-\eta ,2\nu +2,2\rho ),
\end{eqnarray}
where
\begin{eqnarray}
 \rho  = \sqrt{{-2mE \over \hbar ^2}} r \ \ \  {\rm and} \ \ \
   \eta  = {e^2 \over \hbar} \sqrt{m \over -2E}.
   			\label{eq:eta}
\end{eqnarray}
The  mcf in the $r$ coordinate space, $W_r$, for the negative energy state
is thus given by
\begin{eqnarray}
 W_r(r,E,\nu ) &=& \arctan \left[ {\pi \Gamma (b-a) \over
  \Gamma (1-a)\Gamma (b)} {M(a,b,2\rho ) \over V(a,b,2\rho )}
                              \right].  \label{eq:WrEnu}
\end{eqnarray}
The asymptotic forms of $W_r$ at $r = 0$ or $r = \infty$ are
 summarized in Appendix.

 The mcf for the system is given by Eq.~\eqref{eq:mcfSch}
  with expression (\ref{eq:WrEnu}) in place of (\ref{eq:Wr}).
 The  trajectory of an electron in a bound `mode' $(E,\nu ,\mu )$
  should be given by Eqs.~\eqref{eq:MotionS}
   with the same replacement of the variables and parameters
 as mentioned for the scattering state.

 A computer calculation shows that these derivatives
 $ \partial W_r/\partial E$, and $\partial W_r/  \partial \nu $,
 are monotonic with respect to $r$.
  The variation of the derivative $\partial W_r/  \partial \nu $
   between $r = 0$ and $\infty $ cancels out that of the derivative
  $ \partial W_{\theta} /  \partial \nu $ between
  $ \theta  = 0$ and $\pi $.
  It is thus found that the electron moves between $r = 0$ and
 $\infty$, accompanying with a translation between $\theta  = 0$
  and $\pi $.

The trajectory of an electron in a bound state is given as follows.
By assuming the origin to correspond to the nearest point of classical
orbit to the center of the potential and to be the starting point
 of the trajectory,
 the mcf for the motion from $r = 0$ to $r = \infty$ is given by
\begin{eqnarray}
  W = W_r + W_{\theta} + W_{\phi},
\end{eqnarray}
where $W_{\phi} = \mu \phi$, and $W_{\theta}$ or $W_r$ is given by
 Eq.~(\ref{eq:Wtheta}) or (\ref{eq:WrEnu}), respectively.
 The equations of motion are, from (\ref{eq:MotionS}),
\begin{eqnarray}
 \hbar \frac{\partial}{\partial E} W_r - t &=&
 \hbar \frac{\partial}{\partial E} W_r(0) - t_0 = - t_0, \\
 \frac{\partial}{\partial \nu} \left( W_r + W_{\theta} \right)
 &=&  \frac{\partial}{\partial \nu} \left( W_r(0) + W_{\theta}(0) \right)
  = 0,		\\
  \frac{\partial}{\partial \mu} \left( W_{\theta} + W_{\phi} \right)
 &=& \frac{\partial}{\partial \mu} W_{\theta}(0) + \phi_0
  = \pi + \phi_0,
\end{eqnarray}
where $\phi_0$ is the value of $\phi$ at the initial time $t_0$.
 The mcf for the successive motion from $r = \infty$ to  $r = 0$ is
\begin{eqnarray}
  W = - W_r - W_{\theta} + 2W_r(\infty) + 2W_{\theta}(\pi) + W_{\phi}.
\end{eqnarray}
 The equations of motion are
\begin{eqnarray}
  t &=& t_0 - \frac{\hbar \pi \eta}{E}
  - \hbar \frac{\partial}{\partial E} W_r, \,
 \frac{\partial}{\partial \nu} \left( W_r + W_{\theta} \right)
    = 0, \,
  \phi = \frac{\partial}{\partial \mu} W_{\theta} + \pi + \phi_0,
\end{eqnarray}
 where use has been made of $\partial_{\nu} W_r(\infty)
 = - \partial_{\nu}W_{\theta}(\pi) = - \pi$, by  (\ref{eq:Wtheta0pi}),
 and (\ref{eq:Wrinfty}) in Appendix.

  The period of one round trip is found to be
 $ - \hbar\pi\eta/(-E) = 2\pi e^2 \sqrt{m/(-2E)^3}$, which is just
 equal to  that obtained  for the classical orbit with the same energy.
    This suggests that the mcf, (\ref{eq:WrEnu}) with (\ref{eq:Wtheta}),
   is the right one and the mcf could be determined uniquely
     for the system with the mode parameters.

\begin{figure}[htbp]
\begin{center}
  \includegraphics[width=9.5cm]{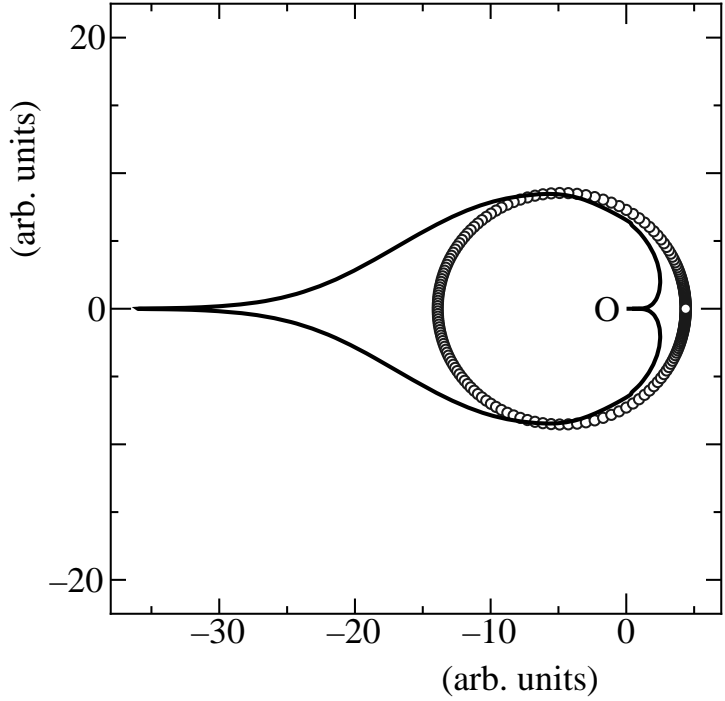}
\end{center}
\caption{ A trajectory of an electron of the hydrogen atom 
in the bound `mode' with $\eta  = 3.3, \nu  = 2.2,$ and 
$\mu  = 1.1$ (solid line) 
and the corresponding classical orbit with $l = 2.7$ (circles).
}
\label{fig:Coulbd}
\end{figure}

 A trajectory of an electron in a bound state thus determined
 is illustrated for the `mode' with parameters
 $\eta  = 3, \nu  = 2, \mu  = 1,$ and $\phi_0 = 0$ at $t=t_0$
 in Fig.~\ref{fig:Coulbd}.
Both of the trajectory or the corresponding classical orbit are
 projected on a plane.
  It could be recognized that the corresponding classical orbit
 resembles more to the trajectory
 as the mode parameters $\eta $ and $\nu $ become larger.
  This shows the correspondence principle.

The requirement that the associated wave (\ref{eq:PhirEnm}) with the round
 trip motion of an electron should be unique everywhere leads to that
  each component of the mcf $W_r, W_{\theta} $ or $ W_{\phi} $ must
 be unique at any point except multiples of $2\pi$ and parameters $\eta,\,
     \nu $ and $\mu $ be integral numbers.
 It gives rise to the usual eigenvalues of the energy for
  the eigenstates.
The wave functions associated with the motion in the $r$ and $\theta $
 coordinates with $\eta, \, \nu $ and $\mu $ integral are composed
  of going and returning waves, Eqs.~(\ref{eq:xjbj})
   and (\ref{eq:xjaj}) with both $-2W_r(0) + 2W_r(\infty)$ and
     $-2W_{\theta}(0) + 2W_{\theta}(\pi)$  being a multiple
 of $2\pi $, or~\cite{KellerRub,Keller}
\begin{eqnarray}
  \sqrt{P_r} e^{iW_r} - \sqrt{P_r} e^{-iW_r}, \,
  {\rm and} \,
  \sqrt{P_{\theta}} e^{iW_{\theta}}
	        - \sqrt{P_{\theta}} e^{-iW_{\theta}}.
\end{eqnarray}
These are  regular for $0 \le r \le \infty$ and
  $ 0 \le \theta \le \pi $, respectively, and equivalent to
  the usual stationary state wave functions except numerical factors.

\section{Conclusion}\label{sec:Concl}
 A trajectory of the electron in an attractive Coulomb potential
 has been shown.
 The  trajectory  resembles well the orbit
 of the corresponding state in classical mechanics.
 It runs also through a tunneling region.
 The trajectory has been derived from the mode characteristic function
 (mcf) which is an extenstion of the Hamilton characteristic function.
  The period of the round trip motion of the electron in the bound state
    is the same as that of the corresponding classical motion
   with the same mode parameters.
 The accordance of the characteristics  between classical and
  quantum mechanics suggests the validity of the mcf and
  the dynamics defined.

It is to be noted that all solutions of the wave equation including
 waves diverging at singular points of the wave equation are necessary
 to derive the mcf.
This suggests the role of all the solutions of the wave equation.
It has been found that if the wave equation is decomposed into
 a set of single-variables $r$, $\theta$ and $\phi$,
 the resulting differential equations  of 2nd order
 have each mcf uniquely, which should lead to
   a mode trajectory of an electron with some dynamical assumptions.

The wave function finite everywhere for any scattering or bound state
 is made by superposing the traveling waves associated
 with the electron motion.
 It is not always continuous for any bound state.
 The requirement that the wave function be continuous or uniqely
 determined at any point of the  coordinate is equivalent to that
 the difference between the mcf's at endpoints be a multiple of $\pi$.
 It leads to that the mode parameters are integral numbers and
 the eigenvalus of the physical quantities become discrete.

The discussion on the cross section and the flux suggests that
the statistical nature of the wave theory stems from the fact
that the wave consists of waves associated with the particle
motion with each mode parameter which is causally determined.
Not all the characteristics of statistical nature of the wave
mechanics have been examined but this may be one of the most
important characteristics.

These results indicate the consistent
 existence of the trajectory in wave mechanics and  suggest
  a significance about the relation between a particle motion
   and a traveling wave function as an extension of the de Broglie
  postulate.

\section*{}%
 The author would like to express his sincere thanks
 to Drs.  S. Nakamura, Y. Takano and T. Okabayashi
  for their continual encouragement during the long course of the work.

\numberwithin{equation}{section}
\appendix\label{App:A}
\section{Asymptotic form of $W_r(a, b, \rho)$ }
 Asymptotic forms of $W_r(a, b, \rho)$ for negative energy states
and its derivatives with respect to $E$  and $\nu$ are summarized.
  It holds from expression~\eqref{eq:WrEnu} that, by putting
 $F_r = \tan W_r$,
\begin{eqnarray}
  {\partial W_r \over \partial E} &=& - {  F_r \over 2E(1 + F_r^2)}
   \Bigg[ \eta \{ \psi(b-a) - \psi(1-a) \} \nonumber \\
   & & - {1 \over M} \left( {\eta \partial M \over \partial a}
  + {\rho \partial M \over \partial \rho } \right)
    + {1 \over V} \left( {\eta \partial V \over \partial a}
     + {\rho \partial V \over \partial \rho } \right) \Bigg], \\
 {\partial W_r \over \partial \nu } &=& {  F_r \over 1 + F_r^2}
  \Bigg[ \psi(b-a) + \psi(1-a) - 2\psi(b) \nonumber \\
   & & + {1 \over M} \left( {\partial M \over \partial a}
     + 2{\partial M \over \partial b} \right) - {1 \over V}
     \left( {\partial V \over \partial a}
       + 2{\partial V \over \partial b} \right) \Bigg].
\end{eqnarray}
As $r$ tends to $0$, it can be written for $b \ge 1$~\cite{AbraSteg,Slater}
\begin{eqnarray}
 M(a,b,2\rho ) &\approx&  1 + {2a \over b}\rho, \\
 V(a,b,2\rho ) &\approx&  \Gamma (b-1) \left[(2\rho )^{1-b} - 1 \right].
\end{eqnarray}
By these approximations, it can be obtained that for $b \ge 1$
\begin{eqnarray}
 W_r(0) = 0, \quad
 {\partial W_r \over \partial E}(0) = 0,  \quad
  {\partial W_r \over \partial \nu }(0) = 0.
\end{eqnarray}
 Asymptotic forms for $r$ large with $a$ and $b$ fixed
  are~\cite{AbraSteg,Slater}
\begin{eqnarray}
 M(a,b,2\rho ) &\approx&  e^{2\rho} (2\rho )^{a-b}
          {\Gamma(b) \over \Gamma(a)}, \\
 V(a,b,2\rho ) &\approx&  - e^{2\rho} (2\rho )^{a-b}
          \Gamma(b - a) \cos \pi a.
\end{eqnarray}

These give rise to
\begin{eqnarray}
 W_r(\infty) = - \pi a, \quad
 {\partial W_r \over \partial E}(\infty )
 = - {\pi \eta  \over 2E}, \quad
 {\partial W_r \over \partial \nu }(\infty ) = -\pi.
			\label{eq:Wrinfty}
\end{eqnarray}
 It is found that the variation of $\partial W_r/\partial E$
  between $r = 0$ and $\infty $ is $\pi \eta /(-2E)$.

\end{document}